\documentclass[conference,table,xcdraw]{IEEEtran}
\usepackage{fancyhdr}
\IEEEoverridecommandlockouts
\usepackage[utf8]{inputenc}
\usepackage{verbatim}
\usepackage{listings}
\usepackage{algorithm}
\usepackage{svg}
\usepackage{epstopdf}
\usepackage{booktabs} 
\usepackage{tabularx}
\usepackage{multirow}
\usepackage{makecell}
\usepackage{setspace}
\usepackage{amsmath,amssymb,amsfonts,amsthm}
\usepackage{graphicx}
\usepackage{color,soul}
\usepackage{textcomp}
\usepackage{latexsym}
\usepackage{courier}
\usepackage{comment}
\usepackage{pifont}
\usepackage{algpseudocode}
\algrenewcommand\textproc{}
\usepackage[all]{nowidow}
\usepackage{array}
\usepackage{pifont}
\newcommand{\cmark}{\ding{51}}%
\newcommand{\xmark}{\ding{55}}%
\usepackage[skip=0pt,font=small]{caption}
\usepackage{bm}
\usepackage{tikz}
\usepackage{cite}
\usepackage{units}
\usepackage{framed}
\usepackage{enumitem}
\usepackage{tikz}
\usepackage{xcolor}
\DeclareRobustCommand*\circled[1]{\tikz[baseline=(char.base)]{
            \node[shape=circle,fill,inner sep=1.3pt] (char) {\textcolor{white}{#1}};}}
\setlist{leftmargin=5.5mm}
\def\BibTeX{{\rm B\kern-.05em{\sc i\kern-.025em b}\kern-.08em
    T\kern-.1667em\lower.7ex\hbox{E}\kern-.125emX}}

\newcommand{\rC}[0]{\rowcolor[HTML]{CCCCCC}}
\newcommand{\hC}[0]{\rowcolor[HTML]{333333}}
\newcommand{\tH}[1]{\multicolumn{1}{c}{\textcolor{white}{#1}}}

\newcommand{\JD}[1]{\textcolor{green}{[JD: #1]}}

\newcommand{\revision}[1]{\textcolor{black}{#1}}
\providecommand{\DIFdel}[1]{} 

\begin{document}

\title{Scaling Distributed Deep Learning Workloads beyond the Memory Capacity with KARMA}


\author{
  \IEEEauthorblockN{
    Mohamed Wahib\IEEEauthorrefmark{1}\IEEEauthorrefmark{4},
    Haoyu Zhang\IEEEauthorrefmark{2},
    Truong Thao Nguyen\IEEEauthorrefmark{1},
    Aleksandr Drozd\IEEEauthorrefmark{4},
    Jens Domke\IEEEauthorrefmark{4},
    Lingqi Zhang\IEEEauthorrefmark{3},\\
    Ryousei Takano\IEEEauthorrefmark{1},
    Satoshi Matsuoka\IEEEauthorrefmark{4}\IEEEauthorrefmark{3}
  }
  \IEEEauthorblockA{
    \small{\parbox{\linewidth}{\centering
      \IEEEauthorrefmark{1}
      National Institute of Advanced Industrial Science and Technology, Japan
      \texttt{\{mohamed.attia,nguyen.truong,takano-ryousei\}@aist.go.jp}
    }}
  }
  \IEEEauthorblockA{
    \small{
      \IEEEauthorrefmark{2}
      miHoYo Inc, (This work was done while the coauthor worked in Tokyo Institute of Technology)
      \texttt{lynkzhang@gmail.com}
    }
  }
  \IEEEauthorblockA{
    \small{
      \IEEEauthorrefmark{3}
      Tokyo Institute of Technology, Tokyo, Japan
      \texttt{zhang.l.ai@m.titech.ac.jp}
    }
  }
  \IEEEauthorblockA{
    \small{
      \IEEEauthorrefmark{4}
      RIKEN Center for Computational Science, Kobe, Japan
      \texttt{\{aleksandr.drozd,jens.domke\}@riken.jp,matsu@acm.org}
    }
  }
}

\maketitle
\thispagestyle{fancy}
\lhead{}
\rhead{}
\chead{}
\lfoot{\footnotesize{
SC20, November 9-19, 2020, Is Everywhere We Are
\newline 978-1-7281-9998-6/20/\$31.00 \copyright 2020 IEEE}}
\rfoot{}
\cfoot{}
\renewcommand{\headrulewidth}{0pt}
\renewcommand{\footrulewidth}{0pt}

\begin{abstract}
The dedicated memory of hardware accelerators can be insufficient to store all weights and/or intermediate states of large deep learning models. Although model parallelism is a viable approach to reduce the memory pressure issue, significant modification of the source code and considerations for algorithms are required. An alternative solution is to use out-of-core methods instead of, or in addition to, data parallelism.

We propose a performance model based on the concurrency analysis of out-of-core training behavior, and derive a strategy that combines layer swapping and redundant recomputing. We achieve an average of 1.52x speedup in six different models over the state-of-the-art out-of-core methods. We also introduce the first method to solve the challenging problem of out-of-core multi-node training by carefully pipelining gradient exchanges and performing the parameter updates on the host. Our data parallel out-of-core solution can outperform complex hybrid model parallelism in training large models, e.g. Megatron-LM and Turning-NLG.
\end{abstract}

\section{Introduction}
\label{sec:intro}
Training Deep Neural Networks (DNNs) is increasingly becoming one of the main HPC workloads. As model and dataset sizes for Deep Learning (DL) become increasingly large, the memory requirement for training Neural Networks (NNs) increases dramatically. Even though the latest generation of Nvidia GPUs have up to \unit[32]{GiB} (V100), this capacity remains a major bottleneck in a lot of the cases~\cite{Gholami:2018:IMB:3210377.3210394}. For example, with a large network such as ResNet-200~\cite{he2016deep}, the local batch-size for training cannot be larger than six ImageNet samples, and in ResNet-1001 the local batch size drops down to two samples. This problem is also a challenge for models that require tens of billions of parameters~\cite{Shoeybi2019MegatronLMTM,Rajbhandari2019ZeROMO}, at which the model will not fit into a single GPU and programmers are forced to employ complex model partitioning methods~\cite{Gholami:2018:IMB:3210377.3210394}.

Distributed training can be used if multiple GPUs are available. Data parallelism is a commonly used scheme in which the model is replicated and the training data is distributed. However, this scheme does not reduce memory pressure from model parameters and activations, forcing users to resort to partitioning a model on several devices (\textit{model parallelism}). For instance, the  Megatron-LM model~\cite{Shoeybi2019MegatronLMTM} has 8.3 billion parameters which need at least $16$ GPUs to fit, assuming \unit[16]{GiB} memory per GPU. In addition, even if a given model is relatively small, there are cases where even a single training sample is too large to be processed on a single GPU. Such cases include high resolution medical or satellite images which can go up to \unit[2]{GiB} per sample~\cite{Mathuriya:2018:CUD:3291656.3291743}. In comparison, the widely used ImageNet dataset~\cite{deng2009imagenet} has images that are smaller than \unit[100]{KiB} per sample (re-sized to $224 \times 224$).

Although model parallelism could be a solution, construction of a cost model and significant modification of the code is needed for every model/dataset/system combination~\cite{Gholami:2018:IMB:3210377.3210394,Naoya:sc19,huang2018gpipe}. Another general solution to this memory capacity problem, that we discuss in this paper, is to use out-of-core methods, without or with redundant recompute, to break the GPU memory limitation~\cite{awan2018oc,shriramdynamic,DBLP:conf/bigdataconf/ItoME17,10.1145/3178487.3178491,Ito2019uPurofilingB,10.1145/3373376.3378505}. 

\revision{The first challenge that KARMA must address is how to first derive an efficient out-of-core strategy that reduces the stall in the execution pipeline, i.e. address the device occupancy bottleneck. Prefetching and data swapping in general, is a well-researched area. That being said, general prefetching and swapping techniques are not efficient for out-of-core DL since they don't provide a comprehensive considerations of the concurrency requirements and occupancy w.r.t. DL workloads~\cite{shriramdynamic,awan2018oc}. The main challenge in deriving an efficient prefetching and swapping strategy is to build a robust model for projecting the minimum required concurrency to keep device utilization as close as possible to maximum. This requires taking into consideration specific features and requirements in DL training: reuse of intermediate results from the forward phase in the backward phase, orchestrating complex pipelines in case of distributed training, non-linear dependency between layers, and memory footprint to compute imbalance (i.e. compute is not linearly correlated to the memory footprint). We propose a performance model to derive a capacity-based out-of-core strategy by the means of assuring a minimum concurrency, i.e., available parallelism, that keeps the device at the highest possible utilization. In addition, we identify and utilize any opportunities at which redundantly recomputing layers reduces the stalls in the pretefching pipeline. More specifically, we generate a schedule to interleave the layers designated for swapping with the recompute of layers that are not swapped.}

\revision{The second challenge that KARMA must address is how to enable multi-GPU training; none of the existing out-of-core methods support multi-GPU, since the layers are split between the GPU and CPU, and hence the weight update becomes complicated. To overcome that, we use a pipeline of overlapped gradient exchanges, by groups of layers, and orchestrate the update of the weights to be executed on the CPU, in a heterogeneous fashion, before swapping the layers in and out of the device for the following iteration.} 

In summary, our paper contributes the following:
\begin{itemize}[leftmargin=2.5mm]
    \item \revision{We propose a performance model based on occupancy analysis to estimate the training time of out-of-core methods.}
    \item Based on our performance mode, we propose \emph{KARMA}, an out-of-core method for training DNNs with the PyTorch framework~\cite{NIPS2019_9015}. KARMA includes a novel approach for interleaving capacity-based layer swapping with redundant recompute to assure peak device occupancy with minimum stalls in the training pipeline. Using KARMA, we achieve an average of 1.52x speedup over the state-of-the-art out-of-core \revision{and recompute} methods. 
    \item We further extend the \revision{capability} of KARMA to support data parallelism in multi-GPU training, i.e. the first out-of-core method to support multi-GPUs, by orchestrating weight updates of swapped-out layers to happen on the CPU side.
    \item We demonstrate how data parallel KARMA \revision{outperforms} the model plus data parallel approaches in training models with billions of parameters, e.g. Megatron-LM~\cite{Shoeybi2019MegatronLMTM} (8.3B parameters) and \revision{Turing-NLG~\cite{Rajbhandari2019ZeROMO} (17B parameters)} trained on $2,048$ GPUs. Finally, we demonstrate cases at which data parallel KARMA is the more cost effective methodology when scaling the mini-batch size above the memory limit.
\end{itemize}
\section{Background and Related Work}
\label{sec:related}
\begin{table*}[bt]
\caption{Limitations and Restrictions of Related Approaches; Label Explanation: \emph{OOC} = Out-of-core, \emph{RECOMP} = Redundant Recomputation, \emph{MP} = Model Parallelism, \emph{N} = Number of Layers, \emph{P} = Number of Model Parameters, \emph{MN} is Multi-node}\label{tb:comparison}
\resizebox{\textwidth}{!}{
\centering\small
\begin{tabular}{|c|c|c|c|c|c|c|c|}
\hline \hC
\tH{\textbf{Name}} & \tH{\textbf{Approach}} & \tH{\textbf{Min.Req. Memory}} & \tH{\textbf{Universal}} & \tH{\textbf{Multi-node}} & \tH{\textbf{Strong Scaling (MN)}} & \tH{\textbf{Fault Tolerance (MN)}} & \tH{\textbf{Ref.}}  \\ \hline\hline
vDNN++ & OOC & None & \xmark & \xmark & N/A & N/A & \cite{shriramdynamic}   \\ \hline  \rC
ooc{\_}cuDNN & OOC & None & \xmark & \xmark & N/A & N/A & \cite{DBLP:conf/bigdataconf/ItoME17}    \\ \hline
 Gradient Checkpoint & RECOMP & $\mathcal{O}(\sqrt{N})$ & \cmark & \cmark & \xmark & \cmark & \cite{Chen2016TrainingDN}    \\ \hline  \rC
SuperNeurons & OOC \& RECOMP & $\mathcal{O}(\sqrt{N})$ & \xmark & \xmark & N/A & N/A & \cite{10.1145/3178487.3178491}  \\ \hline
PoocH & OOC \& RECOMP & $\mathcal{O}(\sqrt{N})$ & \xmark & \xmark & N/A & N/A & \cite{Ito2019uPurofilingB}    \\ \hline  \rC
\revision{Graph Partitioning} & \revision{Implicit MP} & \revision{None} & \revision{\cmark} & \revision{\xmark} & \revision{\xmark} & \revision{\xmark} & \revision{\cite{10.5555/3305890.3305932}}   \\ \hline
FlexFlow & Explicit MP & $\mathcal{O}(\sqrt{P})$ & \xmark & \cmark & \cmark & \xmark & \cite{jia2018beyond}   \\ \hline\hline
\textbf{KARMA (This work)} & OOC \& RECOMP & None & \cmark & \cmark & \cmark & \cmark &  \\ \hline
\end{tabular}}
\end{table*}

Deep neural networks are made up of a network of neurons (nodes) that are organized in layers to form a model. A DNN is trained iteratively by updating the weights of connections between nodes in order to reduce the prediction error of labeled datasets.
A DNN is trained on a dataset of samples to calculate the model weights for which the loss function is minimized. DNNs are trained in two stages. First, during the forward phase, the samples pass through the entire network. Next, in the subsequent backward phase (back propagation), the gradients are computed and weights are updated. Commonly, the samples are randomly chosen from the dataset in batches (known as mini-batch). The training process is performed on the batches of samples, in an iterative fashion, by using an optimization algorithm such as the Stochastic Gradient Descent (SGD). Training with the entire set of training samples is typically repeated, in what is called epochs, until the model convergences to a desired accuracy.

\subsection{Related Work}
Speeding up DNNs is extensively researched, but few works address the memory barrier in distributed training. We do not consider comparisons to basic virtual memory methods (i.e. CUDA Unified Memory) since several works report they perform less favorably than dedicated methods~\cite{shriramdynamic,awan2018oc}.
\subsubsection{\revision{{\bf Out-of-core (Virtualization)}}}
vDNN++~\cite{shriramdynamic} acts as a runtime memory manager that virtualizes the memory usage of DNNs in order to enable simultaneous training of DNNs with both GPU and CPU memories when the DNN cannot fit into GPU memory. Synchronization of the computation and swap-in/out of data at the end of each layer can cause inefficiencies to some extent in vDNN++. The library ooc{\_}cuDNN~\cite{DBLP:conf/bigdataconf/ItoME17} is an extension of Nvidia's DNN library(cuDNN). It uses an out-of-core approach to apply cuDNN-compatible operators even when a layer exceeds GPU memory capacity. In ooc{\_}cuDNN, the feature map of a single layer can be divided over any of the three dimensions: batch, channels, and filters. This means that the swapping of tensors in ooc{\_}cuDNN is limited to the scope of a single layer, i.e., no prefetching is applied.

\subsubsection{\revision{{\bf Gradient Checkpointing (Recompute)}}}
Gradient checkpointing\revision{~\cite{Chen2016TrainingDN,akiba,mlsys2020_196}} is a method that enables fitting larger models into memory by redundantly recomputing activations from checkpoints in the back propagation (at the cost of increased compute time). 

\revision{Since gradient checkpointing, by definition, trades performance for memory capacity, it could not be used to improve performance in distributed training (unlike out-of-core, as will be shown later).} In addition, gradient checkpointing has a lower bound on the memory consumption for a model of $N$ layers of $\mathcal{O}(\sqrt{N})$ \revision{(or $\mathcal{O}(log_{2}N)$ under a special recursive scheme, so may not be used commonly~\cite{Chen2016TrainingDN})}. Hence, it might enable training larger models, yet is still bounded by a memory requirement, unlike out-of-core solutions that in theory have no bounds. 

\subsubsection{{\bf \revision{Out-of-core Plus Recompute:}}, {SuperNeurons, Capuchin, and PoocH}}
SuperNeurons~\cite{10.1145/3178487.3178491} uses a simple method to combine swapping and recomputing based on the target layer, e.g. activations of convolution layers are swapped out while batch normalization layers are recomputed. However, this does not take the actual execution time and memory usage during training into account. PoocH~\cite{Ito2019uPurofilingB} extends a DL framework to enable out-of-core DNN training by mixing layer swapping and recomputing based on a schedule. PoocH decides on the schedule based on the classification of operators in a given model, yet the target layers for swapping or recomputing are determined based on runtime profiling. \revision{Capuchin~\cite{10.1145/3373376.3378505} uses dynamic tracking of tensor access patterns at runtime. It combines out-of-core with recompute to achieves the same amount of footprint reduction as the sole out-of-core approach but with better performance (7\%).}

\revision{It is important to emphasize that all out-of-core methods do not support multi-GPU, to the authors knowledge. That is since the weight update regime for multi-GPUs conflicts with swapping-out layers to CPU. With the massive increase in the resources required for SoTA models (specially models used in NLP), an out-of-core solution supporting multi-GPU training is beneficial for democratizing SoTA NLP research.} 

\subsubsection{{\bf Model Parallelism}}
Model parallelism is increasingly gaining traction as a method to avoid the memory capacity limitation by splitting the model over different GPUs~\cite{gholami2017integrated,jia2018beyond,Naoya:sc19,huang2018gpipe,jia2018exploring}. FlexFlow~\cite{jia2018beyond} notably enables model parallelism by splitting any of the dimensions for any type of DNNs. Model parallelism approaches are invasive and require alteration of how the model is implemented---on a case by case bases---depending on how the model tensors are split. Furthermore, using model parallelism ties the minimum number of GPUs to be used to the model size. For instance, the Megatron-LM model~\cite{Shoeybi2019MegatronLMTM} requires a minimum of $16$ GPUs of \unit[16]{GiB} memory capacity. Even solutions that are highly optimized for reducing the number of GPUs for extremely large models still require tens or hundreds of GPUs~\cite{Rajbhandari2019ZeROMO}.
In summary, model parallelism approaches are complex, intrusive, and enforce a non-trivial bound on minimum number of GPUs for large models.

\revision{Implicit model parallelism methods partition the dataflow graph over the GPUs~\cite{10.5555/3305890.3305932,wang2019supporting,10.1145/3154842.3154843}. One limitation of this pure DAG approaches is the complexity in scaling to multi-nodes, hence DAG solution so far have been limited to a single node. In addition scheduling DAGs is an NP-hard problem with $O(V^2)$ complexity, hence work based on that approach report significant time and resource overhead\cite{10.5555/3305890.3305932,10.1145/3154842.3154843}.} 

\subsection{Summary of Limitations and Issues in Related Approaches}
We highlight the limitations and issues in existing approaches (see Table~\ref{tb:comparison} for summary): 
\begin{itemize}[leftmargin=2.5mm]
\item All prior out-of-core methods are limited to a single GPU. However, unless out-of-core solutions support data parallel multi-node training, they will not be considered a practical approach for training large models and/or datasets. 
\item Using recompute with data parallel training always adds an overhead. Recomputing layers is used with some of the out-of-core methods to further relax the memory limitation, \revision{yet in most cases they add an overhead}. Contrarily, we use recomputing to reduce the runtime by reducing the stalls in the pipeline.
\item Both gradient checkpointing and model parallelism enforce a lower bound on required memory, but this can be an issue when the lower bound is itself higher than the memory capacity per device.
\item Fault tolerance is a concerning issue with single-GPU out-of-core methods and model parallel methods. On the contrary, out-of-core data parallelism (i.e. our KARMA methodology) could potentially adapt to faults by either relaunching with a smaller worker pool~\cite{pmlr-v97-qiao19a} or shrinking the worker pool~\cite{10.1145/3127024.3127037}.
\end{itemize}
\section{KARMA: Out-of-core Distributed Training of Deep Neural Networks}
\label{sec:model}
\subsection{\revision{Overview of KARMA}}
\begin{figure}[t]
\centering
\includegraphics[width=\linewidth,keepaspectratio]{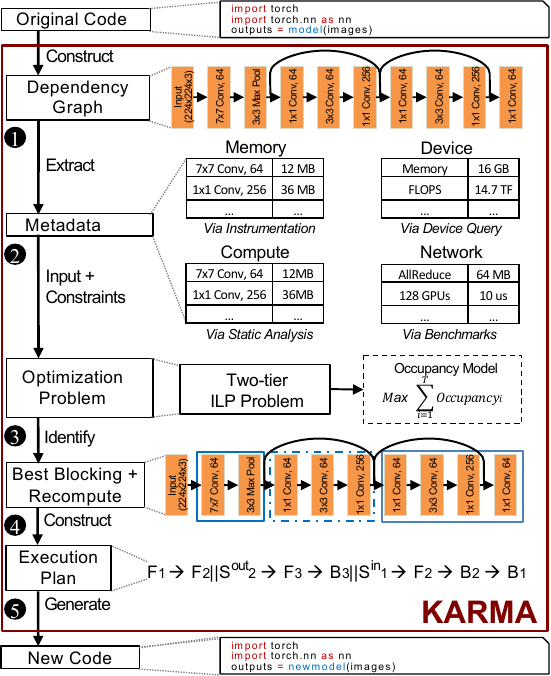}
\caption{\revision{Overview of KARMA's workflow}}
\label{fig:overview}    
\end{figure}

\revision{KARMA enables distributed training of DNNs beyond memory capacity. As shown in the overview in Figure~\ref{fig:overview}, KARMA builds a dependency graph of the model and conducts various examinations on the model to extract metadata that would be used in the performance model of device occupancy (}\circled{1}~\revision{and} \circled{2}~\revision{in the figure). Next, KARMA splits the layers to groups of blocks that optimize for reducing the total runtime. We reduce the total runtime by formulating a two-tier constrained optimization problem that maximizes the device occupancy, which in-turn requires an efficient strategy to reduce the stalls due to data movement to minimum (}\circled{3}~\revision{and} \circled{4})\revision{. We use a novel scheduling strategy that involves a capacity-based layer swapping policy interleaved with recomputation. Next, KARMA generates an execution plan based on the identified optimal blocking and recompute strategy} \circled{5}\revision{), and finally replaces the original model code with the new one.}

\revision{It is important to note that KARMA supports multi-GPU training. In the case of multi-GPU, steps} \circled{4}~\revision{and} \circled{5} \revision{change to introduce a 5-stage pipelined heterogeneous approach for which the orchestration and execution plan is based on pipelining phased gradient exchanges with CPU-side weight update.}


\subsection{Assumptions \revision{and Restrictions}}
The model in this paper can be generalized to discrete accelerators, and is not limited to GPUs. The work in this paper is based on the following assumptions:
\begin{itemize}[leftmargin=2.5mm]
    \item We distinguish two levels of memory: GPU memory is referred to as \emph{near memory} and host memory is \emph{far memory}.
    \item The interconnect between the two memories is bi-directional (e.g. PCIe or NVLink).
    \item Recomputation, i.e. redundant computation, can be utilized by the performance model.
    \item Swapping plus recompute strategy and scheduling is applied on blocks of layers\footnote{For the reminder of this paper we use the word \emph{block} to describe a set of consecutive layers that are bundled together when they are computed, swapped, and their weights are being updated}.
    \item In most cases, the amount of time required to compute a block is less than the time required to swap-in/out this block. 
    \item \revision{The following families of models are supported: CNNs, Transformer-based architectures, and fully convolutional (e.g. U-Net).}
\end{itemize}
\subsection{Computational Cost of Blocks}
\label{sec:layer}
We base our performance model and swapping strategy on the compute and memory requirements of blocks. A multitude of optimizations are used in DL frameworks~\cite{NIPS2019_9015}.
Recent studies of optimizations in DL frameworks show that those optimizations have minimal effect on the aggregate number of operations when combining layers together~\cite{DBLP:journals/jsa/MittalV19}. Few exceptions remain, such as fusing convolution layers with average pooling using summed array tables~\cite{8168154}. Therefore, our proxy of the computational cost for each block is the aggregate number of arithmetic operations (i.e. FLOPS) for all layers in the block. We then use the concurrency capability of the target architecture to project the throughput of processing the blocks on the target architecture.

In the following, we list the compute requirements for layers that most commonly appear in DNNs\footnote{Less common layers are outside the scope of this paper. However, our performance model is generic: it allows adding new layers, if required.}:

\subsubsection{{\bf Convolution Layer}}
For the kernel size $K \cdot K$ and  $C$ channels,
a convolution layer needs $K \cdot K \cdot C_i$ multiply and add operations. The number of operations for direct convolution of one sample is:
\[|Y_i| \cdot K \cdot K \cdot C_i = W_{i+1} \cdot H_{i+1} \cdot C_{i+1} \cdot K \cdot K \cdot C_i\]. Whenever other convolution algorithms are used in cuDNN to compute the layer, e.g. GEMM-based, we adjust the number of operations accordingly based on the algorithm type.

\subsubsection{{\bf ReLu Layer}}
The output of a ReLu layer is $y_i$ = max(0,$x_i$). This requires $|Y_i|$ comparison operations. 

\subsubsection{{\bf Pooling Layer}}
The total number of operations is:
\[|Y_i| \cdot K \cdot K \cdot C_i \cdot c = W_{i+1} \cdot H_{i+1} \cdot C_{i+1} \cdot K \cdot K \cdot C_i\] where $c$ is a multiplier adjusted to the number of operations based on the pooling type (i.e. max vs. average) 

\subsubsection{{\bf Batch Normalization Layer}}
Computing the mini-batch mean requires $|B|$ operations and the mini-batch variance requires $2\cdot|B|$ operations. The normalization and the scaling/shifting requires $4\cdot|X_i|$ and $2\cdot|Y_i|$, respectively. The total number of operations is: $3\cdot|B| + 4\cdot|X_i| + 2\cdot|Y_i|$.

\subsubsection{{\bf Recurrent Neural Network Layer}}
Taking LSTM\cite{Hochreiter1997} as the most common type of layer in RNN, four tensors that form the output of the layer are combined together as input to form the state of the cell by five operations: the total number of operations is $20\cdot|Y_i|$. It is important to note that we adapt the number of operations to the specific RNN variant we use in the model we test. For instance, in attention models~\cite{DBLP:journals/corr/BahdanauCB14} used in machine translation, the decoder RNN additionally receives a weighted sum of encoder hidden states (attention mechanism), rather than just the last hidden state of the encoding stage.

\subsubsection{{\bf \revision{Self-Attention Layer}}}
\revision{Given an input with $Q$ queries’, $K$ keys’, and $V$ values’ matrices, for queries and keys of $d_k$ dimensions and values of $d_v$ dimensions and with a dot-product as a compatibility function $(q.k=\sum_{i=1}^{d_k}q_{i}k_{i})$~\cite{DBLP:journals/corr/BahdanauCB14}, the total operations required are $4d_k^3+d_k^2+2|d_k|$, where \textit{Attention}$(Q, K, V)$ = \textit{Softmax}$(\frac{QK^T}{\sqrt{d_k}})V$.}

\subsubsection{{\bf Fully-Connected Layer}}
Since $y_j \gets f(\sum (x_iw_{i,j}) + bias_{i,j})$, $|WT_i| = |X_i| \times |Y_{i}|$, and $|Y_i| = C_i$, the total number of operations is: $|WT_i|$
\subsubsection{{\bf Softmax Layer}}
Since the softmax layer normalizes its input to a probability distribution, the total number of operations is $2|X_i|$.
\subsubsection{{\bf Other}}
We do not list here the number of operations for other layers/operators which we support in our performance model since they can be simply inferred: dropout, tensor rescaling/reshaping, tensor element-wise, and tensor add.
\subsection{Memory Requirements}
\label{sec:mem}
Unlike the computational cost, the memory requirement for the block depends significantly on the optimizations used by the memory manager, such as layer fusion, changing memory layouts, and work space optimizations. Hence, simple aggregation of memory requirements per layer to estimate the memory requirements for a block could be highly inaccurate~\cite{Li2019BenanzaA}. We use an empirical method that relies on measurements done on a specific target hardware. Note that the empirical results gathered through profiling is done offline. Empirically measuring actual memory used by blocks and layers is not a straightforward task. Most DL frameworks, including PyTorch (which we use in this paper), use a caching memory allocator or manager to speed up memory allocations. As a result, the values shown in CUDA profilers usually don’t reflect the true memory usage. In addition, only differentiable training-related variables and tensors are retained per default, i.e. variables that require gradients will persist. Major DL frameworks often provide low-level APIs to monitor their memory allocator at fine granularity. We conduct a set of experiments to gather information on the memory required per layer, and also fused layers commonly occurring together, using PyTorch's $memory\_stats()$ API that monitors the memory occupied by tensors. We use Nvidia's profiling tools to monitor the memory required for non-model data, e.g. contexts created on the GPU. Finally, after profiling once for each model, we break down and analyze memory use patterns per variable type, i.e. inputs, weights, weight gradients, activations, and activation gradients. This way we can project the memory requirements when we increase the mini-batch size without the need to repeat the offline profiling step.

\subsection{Proposed Capacity-based Strategy for Swapping Blocks}
\label{sec:capa}
\subsubsection{\bf {Occupancy Analysis}}
The goal of our model is to assure we maintain the highest possible occupancy\footnote{Occupancy in this context is not to be mixed with the occupancy metric of CUDA, but refers to the relative amount of compute time, see eq. (1).} of the device. Occupancy in this context factors in the cost of stalling when overlapping layer swapping and prefetching. 
%
The occupancy~$\mathbb{O}$ in step~$j$ can be represented, using the device's busy time $T^\text{busy}_{j}$ and ideal time $T^\text{idle}_{j}$, as follows:
\begin{equation}
\mathbb{O}_j := \text{Occupancy}_{j} = \frac{T^\text{busy}_{j}}{T^\text{busy}_{j}+T^\text{idle}_{j}} 
\end{equation}
For sake of simplicity, we assume buffers of variable sizes, where a buffer size is set equal to the total size of the data arrays containing one or more layers. Hence, the number of available buffers $\mathfrak{B}$, capable of holding data arrays for active layers in GPU memory at the current time step $j$, can be a proxy for occupancy at step $j$, which means the occupancy, we seek to maximize, can also be formulated as:
%

\begin{equation}
\mathbb{O}_j \approx   \left\{ \begin{aligned} &\frac{\mathfrak{B}^\text{avail}_j}{\mathfrak{B}^\text{requ}_j}\\
&1,\ \ \text{for}\ \ \mathfrak{B}^\text{avail}_j > \mathfrak{B}^\text{requ}_j
 \end{aligned}
\right.
\end{equation}
To get the number of available buffers $\mathfrak{B}^\text{avail}$, we have to know which buffers are left after processing at the previous step, and which layers have been swapped in during the same time period. So, for $\mathfrak{B}^\text{avail}_1:=\{\text{entire GPU memory}\}$ and $j\in\mathbb{N}$, we can calculate $\mathfrak{B}^\text{avail}$ in step~$j$ as:
\begin{equation}
\mathfrak{B}^\text{avail}_j = \left\{ \begin{aligned}
&\mathfrak{B}^\text{avail}_{j-1} - \sum_{i=1}^{j-1}(\mathfrak{B}^\text{swapped-in}_i - \mathfrak{B}^\text{processed}_i)\\
&0,\ \ \text{for}\ \ \mathfrak{B}^\text{avail}_{j-1} \leq \mathfrak{B}^\text{swapped-in}_{j-1} - \mathfrak{B}^\text{processed}_{j-1} \end{aligned}
\right.
\label{eq:bufavailj}
\end{equation}
where $\mathfrak{B}^\text{swapped-in}_i$ and $\mathfrak{B}^\text{processed}_i$ are the swapped-in and processed buffers at step $i$, respectively. Note that the buffers are released and tagged as available after processing on them has finished. Obviously, if the rate of swap-in grows faster than processing, then the value of $\mathfrak{B}^\text{avail}$ will approach $0$.
As $\mathfrak{B}^\text{avail}$ keeps decreasing,  $\mathfrak{B}^\text{avail}_{j-1}$ could become less than the required $\mathfrak{B}^\text{requ}_{j-1}$, and hence occupancy~$\mathbb{O}$ would then fall towards~$0$.

If processing is faster, which is typically the case, the value will always be bounded by the overall block swap-in throughput $\mathbb{T}^\text{swap-in}$ that can be inferred from the hardware parameters:
\begin{equation}
\mathbb{T}^\text{swap-in} = \min\{\ \mathbb{T}^\text{FM},\mathbb{T}^\text{NM},\mathbb{T}^\text{IC}\ \}
\label{eq:swapinth}
\end{equation}
Where $\mathbb{T}^\text{FM}$, $\mathbb{T}^\text{NM}$, and $\mathbb{T}^\text{IC}$ are the block-adjusted throughput for the far memory (host memory), near memory (device memory), and inter-connection (PCIe/NVlink), respectively. As discussed previously, if swap-in is faster than processing, the memory consumption will finally reach the GPU memory capacity limit, called $\mathfrak{C}_\text{GPU}$ hereafter, and the capability of swap-in will be affected by the memory space left (since one would have to wait for buffers to clear). If there is not enough memory space for swap-in, the number of buffers swapped in at time step $j$ will be limited to $\mathfrak{B}^\text{avail}_{j-1}$.

To control the granularity of the swapping, we divide the layers into disjoint blocks in both the forward and backward propagation phases. The start time of each block $b$ is the start time of processing or swap-in, and for simplicity we assume they are equivalent. The processing time $T_\text{proc}(b)$ defines that both swap-in and processing finished. Hence, blocks swapped-in depend on the available buffers from the previous time step:
\begin{equation}
\begin{split}
\mathfrak{B}^\text{swapped-in}_j = \min\{\ \mathbb{T}^\text{swap-in} \cdot T_\text{proc}(b)\ ,\ \ \mathfrak{B}^\text{avail}_{j-1}\ \}
\end{split}
\label{eq:time_max}
\end{equation}

Accordingly, for active blocks $b$ in time step $j$, the occupancy can be approximated via:
\begin{equation}\small
\mathbb{O}_j \approx \max \bigg\{ \frac{\mathfrak{B}^\text{avail}_j}{\sum\limits_{b}(\mathfrak{B}^\text{processed}_{j}(b) + \mathbb{T}^\text{swap-in}\cdot T_\text{proc}(b))}, 1 \bigg\}\ , 
\end{equation}
where $\mathfrak{B}^\text{avail}_j$ is defined in Equation~\ref{eq:bufavailj}. 

\subsubsection{{\bf Occupancy in Capacity-based Strategy}}
\begin{figure}[tb]
\centering
\includegraphics[width=\linewidth,keepaspectratio]{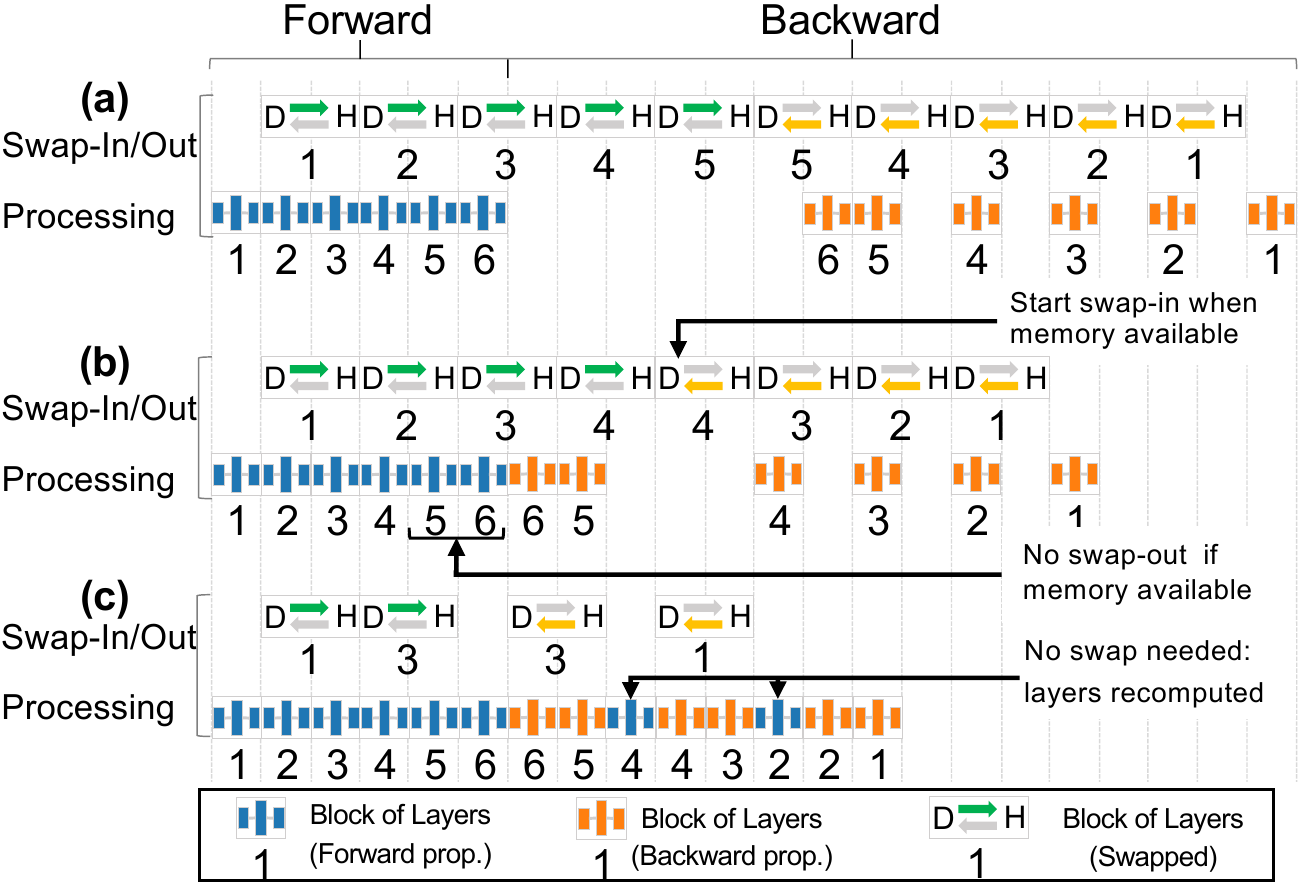}\\
\caption{Different swapping strategies (simplified illustration that assumes the swap-in/out of one block of layers at a time, with swap-in/out taking double the time of the computation). (a) vDNN~\cite{rhu2016vdnn} and ooc{\_}cuDNN~\cite{DBLP:conf/bigdataconf/ItoME17} family of solutions may lead to inefficiency when switching from forward to backward propagation, (b) Proposed capacity-based schedule can reduce the idle time by swapping in ahead of time and avoiding swaps when device capacity allows, (c) Improving capacity-based schedule to further reduce the idle time: swap-in ahead of time interleaved with recomputing following layers}
\label{fig:main}
\end{figure}
In this section we refine the occupancy analysis of the previous section to accommodate our capacity-based strategy. First we take a simple example from the related work, e.g. vDNN~\cite{rhu2016vdnn}, and illustrate the shortcomings of their swap-in/out strategy (shown in Figure \ref{fig:main} (a)). During forward propagation, only swap-out happens. The simple eager strategy used is to swap-out whenever one block's forward step is concluded (including the last block). This can cause performance inefficiency before the backward phase starts. More specifically, the GPU has to wait until the last block is swapped out and then swap it in again before the backward phase can start. During backward propagation, the buffer of block $b-1$ starts swap-in at the same time as block $b$ starts processing. Accordingly, block $b$ will run right after both swap-in and $b+1$ processing is finished. In this case, different layers have different processing times, while we can assume the attributes related to swap-in/out stay the same. If swap-in of $b$ takes more time than processing $b+1$ (which is generally expected), there will be idle times during backward propagation (i.e. stalls), yielding a longer training time.

In this paper we propose an alternative capacity-based strategy for swapping that we explain in following paragraphs. Since the swap-in is only affected by the memory capacity of the GPU, regardless of which block is being processed, we should keep on swapping in as long as we have enough memory space for the buffer. This allows us to keep as much as possible data available for processing at any step (obviously any wait will cause a drop in the average swap-in throughput).

Figure~\ref{fig:main}~(b) shows the proposed capacity-based swap strategy. We get the estimate of the memory consumption from the parameters of the layers in the block by using the empirical method discussed in Section~\ref{sec:mem}. This means even before the forward pass starts, we can know when to stop the swap-out (from block $5$ in this example). When the backward phase starts, the required data would still remain in the GPU memory, and therefore the process can start as soon as the forward stage ends. When one block is processed, the data will be swapped out immediately, in order to make new space for data to be swapped in (i.e. prefetched). In this example, block $4$ can be swapped-in in a very early stage to avoid some GPU stalling caused by data dependency.

Now that we have a clear strategy for swapping data (i.e. capacity-based swap schedule), we can refine the performance model based on this. Let the number of active blocks that can be kept in GPU memory at time step $j$ be $\mathfrak{A}^{blks}_{j}$. At the very beginning of the backward stage, the processing of the blocks will not be impacted by the swap-in. If swap-in is fast enough, all the blocks will be swapped in before the processing of the first block. But if swap-in is relatively slow, the processing may finally catch up with swap-in at a certain time step $\theta$, which satisfies the equation:
\begin{equation}
\label{eq:catch}
\sum_{b\in\mathfrak{A}^{blks}_{\theta}}  T_\text{proc}(b) < \frac{\mathfrak{B}^\text{swapped-in}_{\theta+1}} {\mathbb{T}^\text{swap-in}}
\end{equation}
%
%

Once $\theta$ is reached, the situation becomes the same as discussed in the previous section. In addition, if at time step $\theta$ the inequality in Eq.~\ref{eq:catch} is unsatisfied, that would mean that processing cannot catch-up with data transfer: the whole training process can be at 100\% device occupancy. Now we can derive the refined occupancy:
\begin{equation}
\label{eq:util}
\!\mathbb{O}_j = \left\{
\begin{aligned}
&\mathfrak{B}^\text{avail}_j\!\!\bigg/\!\!\!\sum_{b\in\mathfrak{A}_{blks}}\big(\mathfrak{B}^\text{processed}_{j}(b) + \mathbb{T}^\text{swap-in}\cdot T_\text{proc}(b)\big)\\
&1,\ \text{for}\ T <\theta \text{~where } \theta \text{~occurs when Eq.~\ref{eq:catch} holds}
\end{aligned}
\right.
\end{equation}

The occupancy is the objective function to minimize the training iteration total runtime when using our proposed out-of-core strategy. The cost function for compute is calculated analytically as discussed in Section~\ref{sec:layer}. The swapping time is calculated as $\mathbb{T}^\text{swap-in/out}$ that maximises the occupancy $\mathbb{O}$ in Eq.~\ref{eq:util} for the buffers of each block to be transferred. The same compute and swap models are also used as cost functions for the recompute optimization that is proposed in the next section.
\begin{figure*}[tb]
\centering
\includegraphics[width=\textwidth,keepaspectratio]{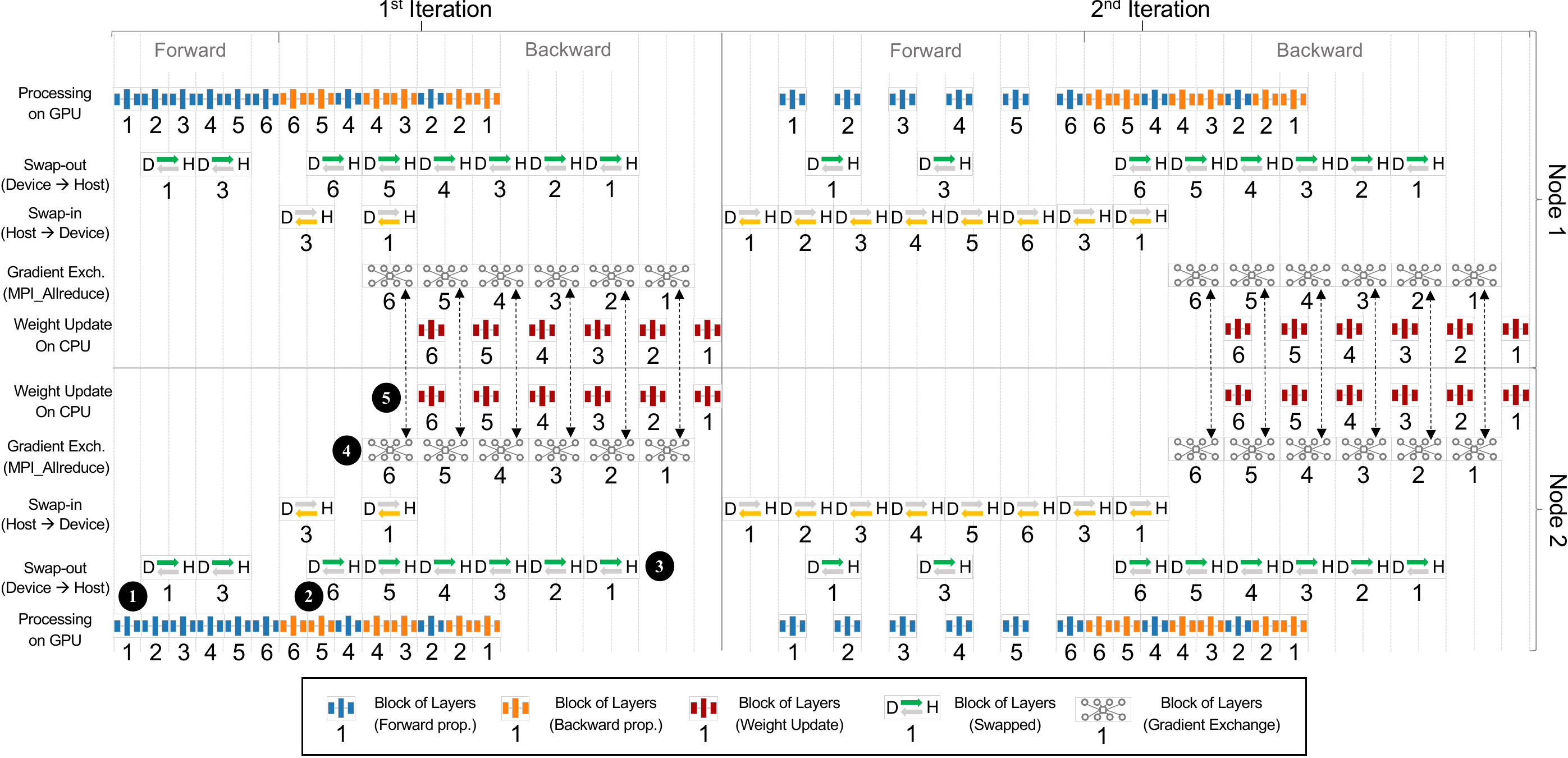}\\
\caption{Using KARMA for data parallel multi-GPU training (an illustrative example of training for two iterations using two nodes). \revision{The capacity-based strategy for layer swapping is interleaved with block recomputation (}\circled{1} \revision{and} \circled{2}\revision{). We overlap the gradient computation with swap-out of blocks to the CPU (}\circled{3}\revision{). We then overlap the gradients exchange (}\circled{4}\revision{) in the backward stage with the weights update (}\circled{5}\revision{) that is done on the CPU side in a heterogeneous manner}, before the blocks are swapped back again to the GPU for starting the following iteration. Iterations after the $2^{nd}$ iteration are similar to the $2^{nd}$ iteration}
\label{fig:multi}
\end{figure*}
\subsection{Interleaving Recompute with the Capacity-based Strategy}
\label{sec:cap_recomp}
The capacity-based strategy discussed in the previous section reduces the swap-in/out overhead by keeping data in near memory as long as there is space, along with a FIFO swapping strategy. \revision{However, this could still lead to gaps in the pipeline at which the processor stalls when waiting for a swap-in to finish (e.g. backward phase of blocks $4$ to $1$ in Figure~\ref{fig:main}~(b))}. The improvement we propose for that method is to interleave the recomputation of the activations of layers in a block with the swap-in of preceding block(s). Those redundantly recomputed layers would have been otherwise swapped-in at the backward stage. The objective of this interleaving is to make sure the computation pipeline remains full and one does not stall due to swapping. The interleaving enables us to overlap compute of the block before the recomputed block with the blocks being swapped-in. As shown in Figure~\ref{fig:main} (c), recomputation of block $5$ could be overlapped with the swap-in of block $3$, and $3$ is overlapped with block $1$. 

\subsubsection{{\bf Dividing Layers to Blocks and Interleaving Recompute: An Optimization Problem}}
In this section we formulate blocking and interleaving as an optimization problem. The execution schedule we define includes a serial sequence of \emph{stages}. Each stage includes a set of blocks, and independent operations on those blocks, i.e. operations on a stage can be overlapped. Deciding on the stages with minimum makespan necessitates the identification of which layers need to be contained in each block, with consideration of the trade off between occupancy and the cost of processing/moving the blocks. In addition, identifying which blocks can be recomputed, rather than swap-in/out to ideally reduce the stalls in the pipeline, is another point to consider. This two-stage optimization problem is a variation of the dynamic clustering problem~\cite{CSPO} (NP-hard), given that the split that defines the first layer in each block is not independent from the split that defines the last layer in the same block (which itself defines the first layer of the following block). We formulate this two-stage optimization as an Integer Linear Programming (ILP) problem to find the boundaries of the blocks that would be swapped, and from that we extract the stages. Next, we further refine the stages to reduce their makespan by identifying the schedule for interleaving recomputation that would minimize the stalls in the pipeline while satisfying the constraints of device memory capacity and dependencies in the pipeline. 

Different alternatives for the optimization were considered, including a multi-objective formulation. However, An important point to note is that: a) pure recompute (i.e. gradient checkpointing~\cite{Chen2016TrainingDN}) and, b) interleaving recompute with swap-in optimize for different objectives. The aim of recompute in gradient checkpointing methods is to relax the memory limitations (with minimal overhead of recomputation). On the other hand, the goal of recompute in this paper is to reduce the makespan by efficiently reducing the stalls in the pipeline. Therefore, doing a two-step optimization allows us to reduce the search space of interleaving permutations by moving from the granularity of layers to the granularity of blocks. Moreover, we assure that the recomputation interleave optimization is independent of the memory constraint by using a separate optimization step for the recompute interleave and moving the memory capacity constraint to the first optimization problem.
\subsubsection{{\bf Problem Formulation}}
For a layer set $L:=\{L_1,\ldots,L_l\}$ split to a set of blocks $B:=\{B_1,\ldots,B_b\}$, finding the execution schedule results in set of stages $S:=\{S_1,\ldots,S_s\}$ having occupancies of $\mathbb{O}(S_i)$, $1 \leq i \leq s$, where each stage includes operations on the blocks to forward compute, backward compute, swap-in, or swap-out (in short: \textit{fw}, \textit{bw}, \textit{in}, and \textit{out}). We have two optimization problems that are satisfied one after the other. First, grouping the layers into blocks that would yield stages of maximum occupancy. Next, refine the stages by identifying blocks to recompute (rather than swap in/out) to reduce the stalls in the pipeline. 

First, the target is to find the $b$-partition $B_1 \cup \ldots \cup B_b$ such that: a) the split of the layers over the blocks is pairwise disjoint and complete, b) operations in a stage are independent and can be overlapped, c) no stage should exceed the device memory capacity, and d) the occupancy of each of the stages $\mathbb{O}(S_i)$ is maximized. Note that for all $1 \leq i \leq s$, $\mathbb{O}(S_j)$ is a dynamic value, i.e., the value is calculated depending on both the make up of the blocks in $S_i$ and the efficiency of the overlap of the operations in the stage. Next, the stages $S$ are refined to new stages $S^{\prime}$ such that a block is recomputed if the recompute time until the next checkpoint is less than the swap-in time of the previous block in the path to the checkpoint. 

The canonical form of the problem is shown in Figure~\ref{fig:karma-opt}. \revision{Constraints ensure the feasibility. Constraints \ref{eq:opt}.1, \ref{eq:opt}.2, and \ref{eq:opt}.3 ensure that the dependencies in the model are not violated. Constraint \ref{eq:opt}.4 enforces the memory capacity limit, and constraint \ref{eq:opt2}.1 marks blocks for recompute rather than swapping, if that reduces the stalls in the execution pipeline.}
\begin{figure}[t]
\small
\begin{spacing}{0.8}
\begin{framed}
\begin{align*}
\text{----~\textbf{Optimization Problem 1}~----}       
\end{align*}
\textit{\hspace{30pt} \textbf{Input}    \hspace{100pt}      \textbf{Output}}
\begin{align*}
& \text{- Layers } L:=L_{1\rightarrow l}
&& \text{- Blocks } B:=B_{1\rightarrow b}\\
\end{align*}
\vspace{-20pt}
\textit{\textbf{Maximize}}
\begin{align}
&\sum_{t=1}^{T} Occupancy_t\label{eq:opt} 
\end{align}
\textit{\textbf{Subject to}} 
\begin{align}
&\sum_{i=1}^{l} \delta_{ij} L_{i} = 1\text{~~~~~~~~~~~~~, }
\forall  j \in  \{1,\ldots,b\} \tag{\ref{eq:opt}.1}\\
&\sum\delta_{ij} = 1 \text{~~~~~~~~~~~~~~~~, } \forall i \forall j \text{~} L_{i} \in B_{j} \tag{\ref{eq:opt}.2}\\[2ex]
&\revision{C_{ij} = 0}\text{~~~~~~~~~~~~~~~~~~~~\revision{,} }
\revision{\forall  \delta_{jm} =1} \revision{\text{~}} \revision{\forall  \delta_{in} =1}\text{\revision{,} } \revision{m<n} \tag{\revision{\ref{eq:opt}.3}}\\
&Mem\left ( t \right )\leq Capacity\text{~~ , } \forall q \in \left \{  1,\ldots,s\right \} \tag{\ref{eq:opt}.4}
\end{align}
\vspace{-10pt}
\\\textit{\textbf{Where}}
\begin{align*}
&\sum Occupancy ~~~~~ \text{Occupancy over time~} T \text{~(Eq.~\ref{eq:util})}\\
&Mem\left (  t\right ) ~~~~~~~~~~~~\text{ Returns device memory required at step ~} t\\
&Capacity ~~~~~~~~~~~~\text{ Is device memory capacity avail. for tensors}\\
&\revision{\delta_{ij}} \revision{\in} \revision{\{}  \revision{0,1} \revision{\}} \text{~~~~~~~~~~~\revision{} }
\revision{\delta_{ij}=1} \text{~\revision{if layer}~} \revision{L_i} \text{~\revision{is in block}~} \revision{B_j}\\
&\revision{C_{ij}} \revision{\in} \revision{\{}  \revision{0,1} \revision{\}} \text{~~~~~~~~~~\revision{} }
\revision{C_{ij}=1} \text{~\revision{if layer}~} \revision{L_i} \text{~\revision{feeds Layer}~} \revision{L_j}\\
\end{align*}
\vspace{-20pt}
\begin{align*}
\text{----~\textbf{Optimization Problem 2}~----}       
\end{align*} 
\textit{\hspace{30pt} \textbf{Input}    \hspace{100pt}      \textbf{Output}}
\begin{align*}
& \text{- Blocks } B_{1\rightarrow b}
&& \text{- New Blocks } B_{1\rightarrow b}^{\prime}\\
\end{align*}
\vspace{-10pt}
\textit{\textbf{Minimize}}
\begin{align}
&\sum_{i=1}^{b} Comp(B_i)\label{eq:opt2}
\end{align}
\textit{\textbf{Subject to}} 
\begin{align}
&\sum_{k}^{d \in \Delta} Comp\left ( B_{k}^{d} \right ) <  \sum_{k}^{d \in \Delta} Swap\left ( B_{k}^{d+1} \right )\text{~,~} k \in \left \{  1,\ldots,b\right \} \tag{\ref{eq:opt2}.1}
\end{align}
\vspace{-10pt}
\\\textit{\textbf{Where}}
\begin{align*}
&\Delta ~~~~~~~~~~~~~~~~~~~~ \text{Set of Blocks until next checkpoint}\\
&Comp\left (  B_m\right ) ~~~~~~\text{ Returns the compute time for block~} B_m\\
&Swap\left (  B_m\right ) ~~~~~~~\text{ Returns the swap time for block~} B_m\\
\end{align*}
\vspace{-30pt}
\end{framed}
\end{spacing}
\caption{Block generation formulated as an optimization problem. A schedule of stages defines both the blocks and the sequence of swaps and recomputations to execute. Each stage includes an overlap of independent operations on different blocks in the stage. Output of optimization problem $1$ is input of optimization problem $2$}
\label{fig:karma-opt}
\end{figure}

\subsubsection{{\bf \revision{Execution Plan Generation}}}
The best feasible solution to the two-step optimization problem in Figure~\ref{fig:karma-opt} is used to generate the execution plan. Algorithm~\ref{alg:sch} shows how we generate the schedule of the stages to execute. The following is an example of the execution plan for the illustrative model in Figure~\ref{fig:main}~(c). For a layer $l$, $F_l$ denotes forward computation, $B_l$ denotes backward computation, $S_l^{in}$ denotes swap-in, and $S_l^{out}$ denotes swap-out. The "$\rightarrow$" symbol denotes the start of the next stage, and "$||$" denotes operations done in parallel:
%
\noindent$F_1 \rightarrow F_2||S_1^{out} \rightarrow F_3 \rightarrow  F_4||S_3^{out} \rightarrow F_5 \rightarrow F_6 \rightarrow B_6||S_3^{in} \rightarrow B_5 \rightarrow F_4 \rightarrow B_4||S_1^{in} \rightarrow B_3 \rightarrow F_2 \rightarrow B_2 \rightarrow B_1$

\subsubsection{{\bf \revision{Support of Non-linear Models in KARMA}}}
\revision{KARMA supports non-linear models at which there are connections between non-consecutive layers: residual networks (e.g ResNet), Transformer-family (e.g. Turing-NLG), and fully convolutional (e.g. U-Net). We inspected the schedules generated by KARMA for different models. For models with affine residual connections, the objective function of the first optimization problem geared the ILP solver towards picking execution plans at which all connections to the layers in a block come from the immediate previous block. Candidate execution plans that have connections that jump over a block would cause a delay in the swap-out and hence be non-optimal solutions.} 

\revision{There are models with non-affine connections, e.g. U-Net has connections from layers in the contracting path to layers in the expansive path. Inspecting the execution plan revealed that the second optimization problem geared the ILP solver towards picking execution plans that changes blocks in the contracting path to recompute, when a layer in the block has an outgoing connection to a layer in the expansive path. Candidate execution plans that would not recompute would be non-optimal solutions since the swapped out blocks in the contracting path would have to be swapped-in prematurely, i.e. before the backward pass.}
\begin{algorithm}[tb]
  \footnotesize
  \caption{Schedule Generation of Stages}
  \label{alg:sch}
  \begin{algorithmic}[1]
  	\Require{Layers: $L$}
    \Ensure{\revision{Stages: $S^\prime = (*_{n}\rightarrow...\rightarrow *_{m})$, $* \in \{F, B, S^{in}, S^{out}\}$}} 
	\State{${S}{\gets}0$, ${TStep}{\gets}0$}\Comment{Initialization}
	\State{${B}{\gets}Opt1()$}\Comment{Blocks defined by $Opt1$ in Figure~\ref{fig:karma-opt}}
	\While {$B\neq\varnothing$}
    	\State{$B^\prime{\gets}getNext(TStep)$}
    	\State{$newStage{\gets}B^\prime$}
    	\State{$S{\gets}add(newStage)$}
    	\State{$B{\gets}B - B\prime$}
    	\State{$TStep{\gets}TStep + 1$}
    \EndWhile
    \State{${B^\prime}{\gets}Opt2()$}\Comment{Blocks inc. recomp. defined by $Opt2$ in Figure~\ref{fig:karma-opt}}
    \State{${S^\prime}{\gets}S$}
	\For {$b\in B^\prime$}
	    \If{$recomp(b) = TRUE$}
	    \State{$s{\gets}getStage(b^{out})$}
	    \State{$S^\prime{\gets}remove(b^{out})$}
	    \State{$S^\prime{\gets}advance(b^{out}_{+})$}\Comment{Advance subsequent swap-out perpetually}
	    \State{$s{\gets}getStage(b^{in})$}
	    \State{$S^\prime{\gets}remove(b^{in})$}
	    \State{$S^\prime{\gets}advance(b^{in}_{+})$}\Comment{Advance subsequent swap-ins perpetually}
	    \State{$newStage{\gets}b^{fw}$}
	    \State{$S^\prime{\gets}insert(newStage)$}\Comment{Insert recomputed block}
	    \EndIf
	\EndFor
  \end{algorithmic}
\end{algorithm}

\subsection{Data Parallel KARMA: An Alternative to Hybrid Data/Model Parallelism}
The rapid increase in model and dataset sizes makes it imperative for KARMA, and out-of-core solutions in general, to support multi-GPU and multi-node training. \revision{However, distributed training is a challenge for all out-of-core solutions, as made evident by the absence of multi-GPU support in all of the previous out-of-core solutions.} 
In single GPU training, the weight update is combined with the backward phase. In typical multi-node or multi-GPU training, where the samples are divided among the GPUs (i.e. data parallelism), the weight update requires a separate step. This separate weight update step follows after computing the gradients and exchanging them among nodes (or GPUs). However, in out-of-core methods, the layers (including their weights) do not entirely reside on the GPU after the end of the backward phase. Since the layers would not entirely fit on the device memory, a trivial workaround is to move the layers entirely to the CPU after the backward phase to apply the weight update there, but this yields an unacceptable performance penalty. 

\revision{Our alternative solution for enabling multi-GPU training requires the expansion of the pipeline of the single GPU (Figure~\ref{fig:main}) from a 2-stage to a 5-stage pipeline. As shown in Figure~\ref{fig:multi}, the three extra stages in the pipeline are as follows:} \emph{First} (\circled{3}), each block is swapped out to the host after computing the gradients for the layers in the block. Note that inter-connect (either PCIe and NVLink) is bidirectional, and therefore swapping out blocks can be overlapped with the swap-in of early blocks in the model to minimize the and stalls introduced by this pipeline stage. \emph{Second} (\circled{4}), rather than exchanging the gradients all at once, we do the AllReduce exchange of the gradients in phases, i.e. finished blocks from the end of the model do the exchange for their gradients without waiting for the other unfinished blocks. The breakdown of the gradient exchange was recently explored as a way to overlap communication and backward computation in distributed synchronous SGD algorithms~\cite{Yamazaki2019YetAA,8638639,Shi2018MGWFBPED}. In this context, however, we do the exchange to the CPU (rather than GPU) since the blocks were swapped out to the CPU before the exchange. \revision{In this work we specifically adopt the layer grouping gradient exchange model by Shi et al.~\cite{Shi2018MGWFBPED}, since it requires only minimal modification to be applied to our approach of grouping layers in blocks.} \emph{Third} (\circled{5}), after the gradients of a block are exchanged, the weights are updated on the CPU. The computational cost for it is larger than on GPU (plus it requires implementing the CPU side update). However, this heterogeneous approach allows us to have a better overlap with negligible overhead. Note that this also alters the stages starting from the second iteration (as Figure~\ref{fig:multi} shows).
\subsection{Integration of KARMA in a Deep Learning Framework}
The model is interpreted during training, since we implement KARMA in PyTorch. For frameworks that use a define-and-run scheme (e.g. TensorFlow), the execution schedule can be encoded statically as a new computational graph, but this is outside of the scope of our work. We extract the forward and backward phases, then after offline profiling we construct and solve the two-stage ILP  optimization problem using the MIDACO solver~\cite{SCHLUTER20092217,midaco}. KARMA then composes the execution schedule from the solution (i.e. stages) and generates a new training script. In the new training script, we use CUDA Unified Memory (UM) technology in addition to a prefetcher. In our implementation, we used $cudaMemPrefetchAsync()$ as a data transfer method. Even though we don't swap data out explicitly, this method allows us to implicitly apply the capacity-based strategy described in Section~\ref{sec:capa}. Since we are trying to overlap data transfer as much as possible, we synchronize before the prefetch to make sure the prefetch will not start too early. Note that PyTorch typically calls the cuDNN library for processing only after all buffers for the layer have been prepared (i.e. swapped in) during backward propagation. We also synchronize after the prefetch to make sure the data is ready to be processed, or we would risk a significant penalty from page faulting.

In an ideal prefetch, the prefetch instruction for block $b$ will be launched to the CUDA stream of the block at the same time that the backward pass on the layers in block $b+1$ start running. In this case, although the backward pass of block $b$ will not even be launched before the data is prefeched properly, it will start much later, so the performance will not be affected by the synchronization.

For multi-node KARMA, we initially relied on Nvidia's NCCL communication library to implement the phased gradient exchanges. Yet, due to NCCL's stability issues that appeared at scale ($>1,000$ GPUs), we instead used the MPI back-end of the PyTorch communicator (\emph{torch.distributed}). Finally, we implemented a stand-alone direct CPU kernel to update the weights of individual blocks. 

\section{Evaluation}
\label{sec:exp}
\begin{table}[tb]
\centering
\caption{Environment Information (ABCI Supercomputer)}\label{tab:env}
\resizebox{\linewidth}{!}{
\begin{tabular}{l|lrr}
\hline
Compute nodes (w/ GPU)& 1,088 (total 4,352 GPUs) & &                 \\ \hline
GPU                  & Nvidia Tesla V100 (SMX2) & x4 & (\unit[16]{GiB})    
\\ 
CPU                  & Intel(R) Xeon(R) Gold 6148 & x2 & (\unit[32]{GiB}$\times$6)
\\ 
CPU-GPU Interconnect & PCI-Express Gen3 x16 & & (\unit[16]{GB/s})     
\\ 
GPU-GPU Interconnect & NVLINK & & (\unit[50]{GB/s})                               
\\ 
System Interconnect              & \unit[100]{Gbps} EDR InfiniBand & x2 & (\unit[12.5]{GB/s})       
\\ \hline
CUDA                 & v10.1 & &                                 \\
cuDNN                & v7.6.1 & &                                 \\ 
PyTorch              & v1.2 & &                                          \\   \hline
\end{tabular}}
\end{table}
\begin{table}[t]
\centering
\caption{Overview of Models and Datasets Used in Experiments}
\footnotesize
\label{tab:models}
\setlength\tabcolsep{3pt} 
\resizebox{\linewidth}{!}{
\begin{tabular}{|l|c|c|c|c|}
\hline
\multicolumn{1}{|c|}{Model} & Dataset             & \# Samples & Parameters & Layers \\ \hline
ResNet-50~\cite{he2016deep}                            &    \multirow{3}{*}{ImageNet~\cite{deng2009imagenet}}     & \multirow{3}{*}{1,280,000}& \textgreater{}~25M      & 50                 \\ \cline{1-1} \cline{4-5} 
VGG16~\cite{DBLP:journals/corr/SimonyanZ14a}                               &                               &            & \textgreater{}~169M     & 38                  \\\cline{1-1} \cline{4-5} 
ResNet-200~\cite{he2016deep}                           &                               &            & \textgreater{}~64M      & 200
\\ \hline
WRN-28-10~\cite{BMVC2016_87}                          & \multirow{2}{*}{CIFAR-10~\cite{Krizhevsky09}}     & \multirow{2}{*}{60,000}              & \textgreater{}~36M      & 28               \\ \cline{1-1} \cline{4-5} 
ResNet-1001~\cite{he2016deep}                                  &            &               & \textgreater{}~10M      & 1001                 \\ \hline
\revision{U-Net~\cite{10.1007/978-3-319-24574-4_28}}      & \revision{ssTEM~\cite{sstem}}     & \revision{30} & \revision{\textgreater{}~31M}                   & \revision{27}                 \\  \hline
Megatron-LM~\cite{Shoeybi2019MegatronLMTM}      & \multirow{2}{*}{OpenWT~\cite{owt}}     & \multirow{2}{*}{7,200,000} & \textgreater{}~8.3B                   & 72                 \\  \cline{1-1} \cline{4-5} 
\revision{Turing-NLG~\cite{Rajbhandari2019ZeROMO}}      &  &  & \revision{\textgreater{}~17B}                   & \revision{78}                 \\  \hline
\end{tabular}}
\end{table}
\renewcommand{\arraystretch}{1.0}
\normalsize
\subsection{Hardware Platform, Models, and Datasets}
The following experiments are all performed on ABCI supercomputer: $8^{th}$ in the Top500 list (November 2019). The specifications are shown in Table~\ref{tab:env}.

Table~\ref{tab:models} lists the models and datasets we use in our experiments and summarizes the relevant features. We use the same pre-processing configurations and hyper-parameter values reported by the cited models, unless mentioned otherwise. The \revision{parallel version of} MIDACO library~\cite{SCHLUTER20092217,midaco} we use to solve the ILP problems converged \revision{in under four minutes} for all of our inputs, and is thus negligible.

\begin{figure*}[t]
\centering
\includegraphics[width=\textwidth]{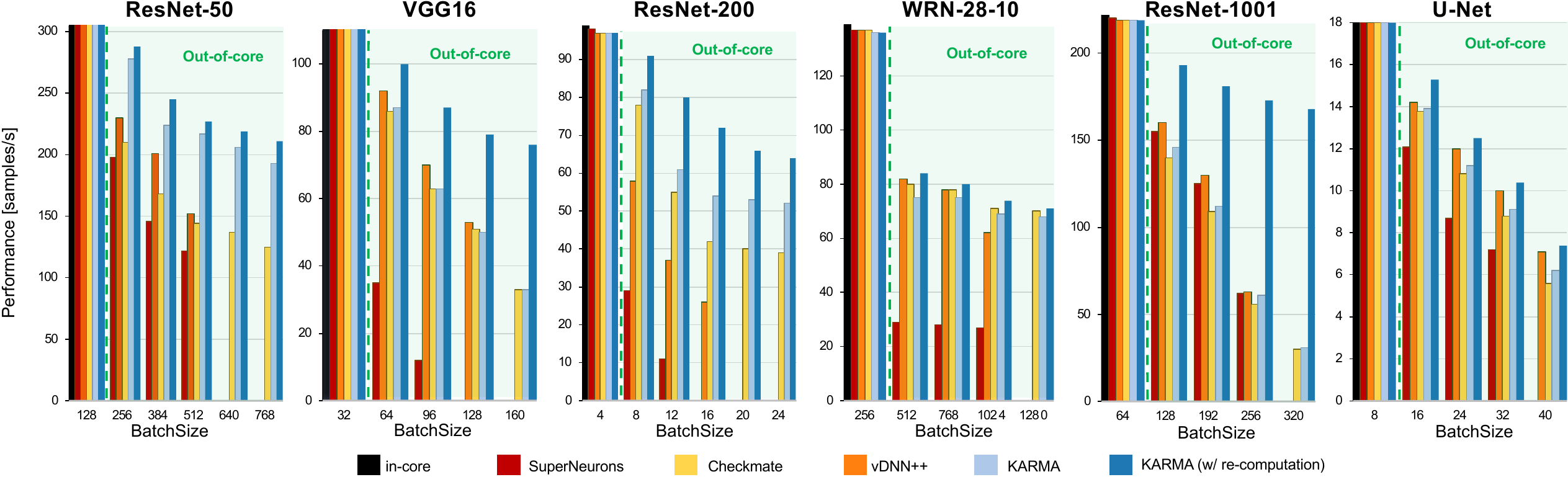} 
\caption{Performance using a V100 SMX2 GPU ($16$GB). For all figures, only the first reported mini-batch size (x-axis) fits in memory}
\label{fig:allperf}
\end{figure*}
\subsection{KARMA for Single GPU}
\subsubsection{{\bf Performance}}
Figure~\ref{fig:allperf} shows the training performance (samples/second) of different batch sizes\footnote{In this paper, \emph{batch size} is synonymous to \emph{mini-batch size}}, for different models. As the batch size grows, the out-of-core effects start to appear. The performance begins to drop after the memory footprint exceeds the GPU memory capacity (starting from the second data point on each x-axis). As discussed earlier, the data movement time should typically be larger than compute time when out-of-core is required, yet the performance does not drop suddenly. Due to our capacity-based strategy, there is a stage where utilization can still be at $100\%$ in the beginning of the backward phase. However, as the input size grows, that duration will become shorter. Finally, the performance converges to be limited by the data swapping throughput.

\revision{In addition to out-of-core methods, we compare with two recompute methods: Checkmate~\cite{mlsys2020_196} which reports state-of-the-art performance and formulates the problem as a constrained optimization problem and solves it using ILP, and b) SuperNeuron~\cite{10.1145/3178487.3178491} (which mixes recompute with out-of-core). KARMA's capacity-based strategy interleaved with recompute outperforms both in all tested models.}
\subsubsection{{\bf Efficiency}}
\begin{figure}[t]
\centering
\includegraphics[width=\linewidth]{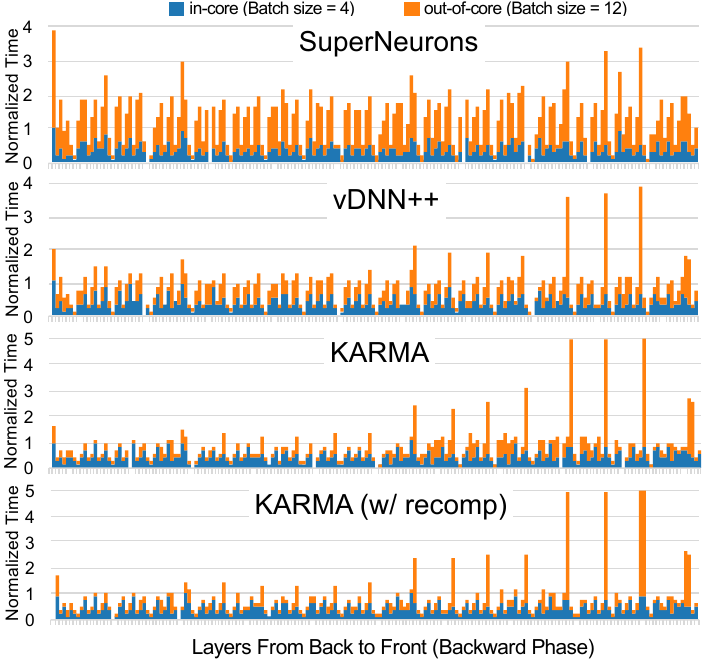} 
\caption{Normalized runtime in the backward phase of RestNet-200: out-of-core run stacked on an in-core run. The orange bars include the runtime normalized to the same batch size in addition to all the stalls from layer swapping and recompute pipeline}
\label{fig:allstall-one-column}
\end{figure}

Figure~\ref{fig:allstall-one-column} gives insight into the effectiveness of combining the capacity-based method with recomputing. The large spikes appearing towards the end are stalls resulting from waiting for the swap-in of the convolution layers which appear early in the model. SuperNeurons's stalls are spread out across the layers due to the inaccuracy of the static cost functions and the designation of swapping vs.~recompute based on the layer type, and not a cost function. KARMA and KARMA w/recompute, and vDNN++ exhibit more spread out stalls. However, vDNN++ suffers from an early large spike. KARMA's capacity-based approach avoids the swapping of late layers in the model, which reduces stalls at the start of the backward phase. Furthermore, vDNN++ includes a significant number of smaller spikes just before the large spikes. KARMA w/recompute is more flat between the large spikes since the interleaved recomputation fills the gaps in the pipeline and reduces the stalls to only a few unavoidable large spikes. 

\begin{figure*}[t]
\centering
\includegraphics[width=\textwidth]{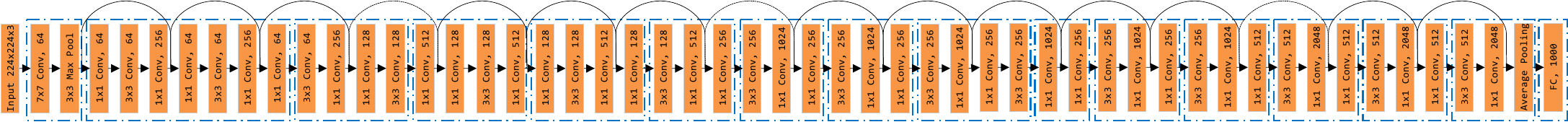} 
\caption{Best blocking found by KARMA for ResNet-50/ImageNet (batch size = $512$) on a V100 SMX2 GPU ($16$GB)}
\label{fig:resnetblocks}
\end{figure*}

Figure~\ref{fig:resnetblocks} shows the best blocking found by KARMA for ResNet-50 on a V100 GPUs. The blocking of layers by KARMA results in a well-balanced overlap of data movement and workload that would reduce the stalling to minimum. Compared to other methods, the execution plan resulting from this blocking reduces the stalling by $43$\% and $37$\% over SuperNeurons and vDNN++, respectively.
\subsection{Data Parallel KARMA (Multi-GPU)}
In this section we discuss the results of data parallel KARMA. Table~\ref{tab:megatron} shows different configurations tested for the state-of-the-art Megatron-LM model (based on GPT-2~\cite{radford2019language}). Megatron-LM configurations generate large models that do not fit in device memory and therefore it was previously implemented using a model/data-parallel hybrid~\cite{Shoeybi2019MegatronLMTM}. This can be a major limitation for practitioners. For instance, the 8.3 billion parameter configuration requires $8$~GPUs with \unit[32]{GiB} memory each for 8-way model parallelism which is prohibitive. We use data parallel KARMA to run Megatron-LM entirely using data parallelism and avoid the complexities of model parallelism. The performance is listed in Table~\ref{tab:megatron}. \revision{The table also shows comparable perplexity in comparison with Megatron-LM on the OpenWebText dataset~\cite{owt} (excluding the last two configurations that are not feasible to train to completion, given the cost of training)}.
\begin{table}[t]
\centering
\caption{Data Parallel KARMA Configurations and Performance for Megatron-LM; Label Explanation: \emph{H} = Hidden Size, \emph{A} = Attention Heads, \emph{L} = Layers, \emph{P} = Parameters, \emph{MP} = Model Parallel, \emph{DP} = Data Parallel, \emph{OCC} = Out-of-core (KARMA), \emph{Perf} = Iter./sec, \revision{and \emph{PPL} = Zero-shot Perplexity for OpenWebText~\cite{owt}}}
\footnotesize
\label{tab:megatron}
\setlength\tabcolsep{3pt} 
\resizebox{\linewidth}{!}{
\begin{tabular}{|c|c|c|c|c|c|c|c|c|c|c|}
\hline
\multirow{2}{*}{H} &
\multirow{2}{*}{A} & 
\multirow{2}{*}{L} & 
\multirow{2}{*}{P} & \multicolumn{4}{c|}{Megatron-LM ($\ddagger$ Num. GPUs)} & \multicolumn{3}{c|}{DP KARMA}\\ \cline{5-11}
& & & & MP$\ddagger$&MP+DP$\ddagger$&Perf&\revision{PPL}&GPUs&Perf&\revision{PPL}\\ \hline
1152& 12& 18& 0.7B& 1& 64&5.8& \revision{13.66}& 32 &2.2& \revision{13.85}\\ \hline
1536& 16& 40& 1.2B& 2& 128&1.6& \revision{10.47}& 64 &0.73& \revision{10.34}\\ \hline
1920& 20& 54& 2.5B& 4& 256&2.9& \revision{8.21}& 128 &1.94& \revision{8.33}\\ \hline
2304& 24& 64& 4.2B& 8& 512&5.0& \revision{N/A}& 256 &3.11& \revision{N/A}\\ \hline
3072& 32& 72& 8.3B& 16& 1024&8.4& \revision{N/A}& 512 &6.3& \revision{N/A}\\ \hline
\end{tabular}}
\end{table}

\begin{figure}[t]
\centering
\includegraphics[width=\linewidth]{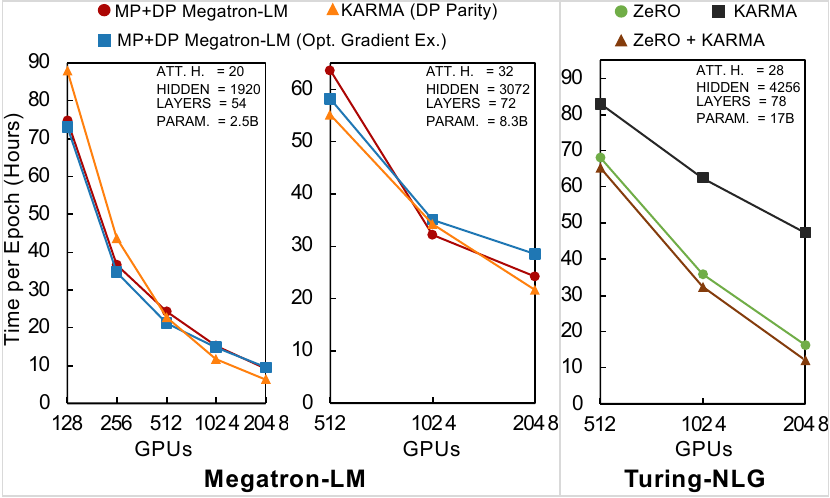} 
\caption{Parallelization performance of two Megatron-LM configurations (Table~\ref{tab:megatron}), \revision{and Turning-NLG}. We compare using the same number of GPUs (parity). Megatron-LM: we compare the original data-/model-parallel hybrid, the original plus our optimized phased gradient exchange, and data parallel KARMA. \revision{Turing-NLG: we compare the hybrid ZeRO reference implementation, data parallel KARMA, and KARMA used on top of data parallel ZeRO. The mini-batch size in KARMA is multiplied by the model parallel factor of the original implementation} }\label{fig:mega}
\end{figure}

We further conduct an experiment at which we use the same number of GPUs (parity comparison\footnote{We use the term \emph{parity} to refer to using a number of GPUs with data parallel KARMA that is equal to the number of GPUs used in the hybrid of model and data parallel}) for the original implementation vs.~data parallel KARMA (Figure~\ref{fig:mega}). To get a fair comparison, we also compare to an optimized version of the original implementation for which we added the phased gradient exchange. Surprisingly, in the parity comparison, the pure data parallel KARMA outperforms the model-/data-parallel hybrid on $2,048$ GPUs. Upon inspection it became clear that increasing the numbers of GPUs also increases the communication cost for the original version. Note that KARMA has fewer iterations (i.e. communication rounds) since it has a larger mini-batch size. \revision{Figure~\ref{fig:mega} also shows results for Turing-NLG~\cite{Rajbhandari2019ZeROMO}, a 17B parameters model with state-of-the-art results in NLP. The Turing-NLG model has $78$ Transformer layers with a hidden size of $4256$ and $28$ attention heads. Turing-NLG is implemented using ZeRO, a memory optimizer that reduces the memory footprint by splitting the model parameters, gradients, and optimizer states among GPUs. Despite the reduction in the memory footprint, large models, such as Turing-NLG, still require a hybrid of model and data parallelism. For the same number of GPUs, KARMA shows less performance than ZeRO, which is expected given the aggressive memory optimizations in ZeRO. However, we also tested KARMA on top of ZeRO, i.e. KARMA enabling the use of the same number of GPUs all in data parallel mode, rather then the model/data parallel hybrid of ZeRO. KARMA+ZeRO gives a speedup of $1.35\times$ over ZeRO for Turing-NLG with up to $2,048$ GPUs.}

It is important to mention that limitation of the system scheduler prevented us from running full epochs of Megatron-LM and \revision{Turing-NLG} at large scale. As a mitigation strategy, we split the epoch into separate runs at which we checkpoint/restart the model state, and add up the individual runtime parts for an entire epoch, ignoring the C/R overhead.

Table~\ref{tab:cost-effective} shows the cost/performance of two data parallel models vs. data parallel KARMA (where KARMA is using fewer GPUs). Interestingly, KARMA can be more cost effective than data parallelism when initially increasing the number of GPUs. This is due to the positive effect of the capacity-based strategy: only an initial small drop in performance when using KARMA to increase the mini-batch size. However, data parallelism becomes more cost effective when further increasing the number of GPUs as the slowdown due to out-of-core start to magnify.  
\renewcommand{\arraystretch}{1.0}

%

\begin{table}[t]
\centering
\caption{Cost/Performance Normalized to \$/P of First Row. Number of Samples/GPU Fixed for Data Parallel (at Max of Memory Capacity). Number of GPUs for Data Parallel KARMA Fixed while Samples/GPU Increase; Label Explanation: \emph{DP} = Data Parallel, and \emph{\$/P} = Cost/Performance (Number of GPUs/training throughput)}
\footnotesize
\label{tab:cost-effective}
\setlength\tabcolsep{3pt} 
\resizebox{\linewidth}{!}{
\begin{tabular}{|r|c|c|c|c|r|c|c|c|c|}
\hline
\multicolumn{5}{|c|}{\small ResNet-50}&\multicolumn{5}{c|}{\small ResNet-200}\\ \hline
\multirow{2}{*}{Batch}&\multicolumn{2}{c|}{DP}&\multicolumn{2}{c|}{DP KARMA}&\multirow{2}{*}{Batch}&\multicolumn{2}{c|}{DP}&\multicolumn{2}{c|}{DP KARMA}\\ \cline{2-5}\cline{7-10}
      & GPUs & \$/P & GPUs & \$/P &  & GPUs & \$/P & GPUs & \$/P \\ \hline
12.8K & 100 &  1 &          100 & 1  & 400  & 100 & 1 &     100 & 1   \\ \hline
25.6K & 200 &  1.040 &      100 &\textbf{1.026} &   800K & 200 & 1.088 & 100 & \textbf{1.06} \\ \hline
38.4K & 300 &  1.051 &  100 & \textbf{1.037} &   1.2K & 300 & 1.093 & 100 & \textbf{1.066}  \\ \hline
51.2K & 400 &  \textbf{1.066} & 100 & 1.378 & 1.6K & 400 &\textbf{1.101} & 100 & 1.263   \\ \hline
  64K & 500 &  \textbf{1.089} & 100 & 1.429 & 2K & 500 & \textbf{1.136} & 100 & 1.371  \\ \hline
76.8K & 600 &  \textbf{1.092} & 100 & 1.483 &  2.4K & 600 & \textbf{1.171} & 100 & 1.412    \\ \hline
\end{tabular}}
\end{table}
\subsection{Accuracy}
The proposed out-of-core strategy has no impact on the accuracy of the model since neither the shape nor the hyper-parameters of the model change. Nonetheless, to assure the correctness of our implementation, we ran selected configurations to convergence and compared with the reported accuracy (same number of epochs and hyper-parameters). ResNet-50/ImageNet with $256$ samples/GPU on a single GPU using KARMA and $16$~GPUs data parallel KARMA achieved $75.9\%$ and $75.8\%$ accuracy, respectively (vs.~$75.9\%$ reported top-1 accuracy for single crop on the validation dataset~\cite{he2016deep}). VGG16/ImageNet with $64$ samples/GPU on a single GPU using KARMA and $32$~GPU data parallel KARMA achieved $72\%$ and $72\%$, respectively (vs.~$71.9\%$ reported top-1 accuracy for single crop on the validation dataset~\cite{DBLP:journals/corr/SimonyanZ14a}). \revision{We also show perplexity values for the experiments on language models in Table~\ref{tab:megatron}}

\subsection{Discussion of Experimental Outcomes}
The results and findings of our experiments, using our novel KARMA methodology, can be summarized as follows: 
\begin{itemize}[leftmargin=2.5mm]
    \item KARMA enables a general increase of the batch size, 2x--6x above the memory limit, with only an average performance degradation of 9\%--37\%, outperforming other approaches.
    \item Our capacity-based strategy and interleaving recomputation significantly reduces the stalls in the swapping pipeline over state-of-the-art methods.
    \item Not only does data parallel KARMA eliminate the need for model parallelism in models larger than memory capacity, it also can yield better runtime performance at large scale when using the same number of GPUs.
    \item In some cases, data parallel KARMA provides better cost/performance compared to utilizing more GPUs in existing data parallel approaches when hitting the memory limit.
\end{itemize}

\section{Conclusion}
GPU memory capacity can be a major bottleneck for DNN training workloads. A general solution to this problem are out-of-core methods. We study the behavior and limitations of out-of-core methods. We propose a strategy that combines a capacity-based layer swapping strategy with interleaved recomputations. Our method outperforms state-of-the-art out-of-core methods, with a 1.52x speedup on average. Finally, we extend our method to multi-nodes by using a phased gradient exchange and CPU-side weight update to construct a complex pipeline. Our multi-node out-of-core method is a viable alternative for training large DL models that currently necessitate complex model parallelism schemes.

\section*{Acknowledgement}
This work was partially supported by JST ACT-X Grant Number JPMJAX190C, Japan and
JST CREST Grant Number JPMJCR19F5, Japan.
\bibliographystyle{IEEEtran}
\bibliography{main}

\end{document}